\documentclass[prd,aps,twocolumn,showpacs,preprintnumbers,%
amsmath,amssymb]{revtex4}
\usepackage{graphicx} 

\def\be{\begin{equation}}

\def\j-{\J_-}

\def\ee{\end{equation}}

\def\be{\begin{equation}}
\def\ee{\end{equation}}

\def\bea{\begin{eqnarray}}
\def\eea{\end{eqnarray}}
\def\bearr{\begin{eqnarray}}
\def\bearrs{\begin{eqnarray*}}
\def\eearr{\end{eqnarray}}
\def\eearrs{\end{eqnarray*}}
\def\barr{\begin{array}}
\def\earr{\end{array}}

\def\non\non{\nonumber}
\def\nn8{\nonumber\\[15pt]}
\def\l{\left}
\def\r{\right}

\begin{document}
\title{Contribution of $U_{e3}$ to geo-neutrino flux}
\author{
Subhendra Mohanty}
\affiliation{ Physical Research Laboratory,
Navrangpura, Ahmedabad - 380 009, India}

\begin{abstract}

We show that a non-zero $U_{e3}$ close to the CHOOZ bound ($Sin^2
2 \theta_{13} \simeq 0.16$ $90\%$ CL) can change the geo-neutrino
flux by 12\%. Geo-neutrino detection in Kamland with a exposure of
$3 \times 10^{32} proton-years$ is sensitive to $Sin^2 2 \theta_{13}$
to the level of $0.2 (1 \sigma)$. For the same exposure a detector close
to
Himalayas can probe $Sin^2 2 \theta_{13}$
down to  $0.15 (1 \sigma)$ due to higher geo-neutrino flux from the
Tibetan plate. 

\end{abstract}

\maketitle


The Kamland measurement
\cite{kamland} of geo-neutrinos \cite{geoneutrinos} has shown the
possibility of using this neutrino source to learn about the earth as 
well as to determine neutrino properties. The first step in using these
measurements would be to determine the geochemical properties of earth
-mainly the Uranium and Thorium distribution \cite{earth} 
using the known bounds on neutrino parameters.
The Bulk Silicate Model of the earth predicts the neutrino flux at Kamioka
$\Phi=(3.7 \pm 0.2) \times 10^6 cm^{-2} s^{-2}$ i.e an accuracy of $5
\% (1 \sigma)$ \cite{bse},\cite{ref}.    
A five kiloton-detector over a period of five years can have enough
statistics to measure the total geo-neutrino events with a statistical 
accuracy of
$5\%$ (at $ 1 \sigma$) \cite{bse}. With this accuracy in theory and
experiment in mind it is worth investigating what neutrino properties
can be tested using the geo-neutrinos.

In this paper we do a three flavour calculation of $\bar \nu_e$ survival
probability assuming a non-zero mixing angle $\theta_{13}$ 
close to the
bounds from CHO0Z
\cite{chooz} and Palo Verde \cite{palo},  $Sin^2 \theta_{13} < 0.16
(0.25) $ at $90\% CL (3
\sigma)$
(assuming $\Delta m_{13} =2.0 \times 10^{-3} eV^2 $ \cite{white}). In a
two
flavour analysis where the neutrino oscillations are determined by the
solar
neutrino mass scale $\Delta_S= 7.3 \times 10^{-5} eV^2$, the oscillation
length at the geo-neutrino energies  is less than $100 Km$. So the
oscillation probability for geoneutrinos which are mostly from larger
distances is well approximated by $Pee=1-0.5
~Sin^2 2 \theta_{12}= 0.57$ (taking $Sin^2 2\theta=0.86$). Full
calculation of the two flavor oscillation probaility reveals that
the energy dependent term changes by less than $(3 \%)$ \cite{ref},
\cite{moh}
as $E_{vis}$  varies in the geo-neutrino
spectrum range $1
MeV-2.5 MeV$. In  three flavor oscillations (which in the geo-neutrino
energy range is given by the formula (\ref{pee}) shown below), the
oscillation lenghth associated with the atmospheric neutrinos mass scale
$\Delta m_a \simeq 2 \times 10^{-3} eV^2$ is larger than the diameter of
the earth so the energy dependent term cannot be averaged $0.5$ and a full
three flavor calculation needs to performed if $Sin^2 2 \theta_{13} \neq
0$.

The three flavour oscillation formula relevant at geo-neutrino energies
and length scales is   
\begin{eqnarray}
P_{ee}(E,L)=1&-&Sin^2 2 \theta_{13} Sin^2 \l(\frac{\Delta_a L}{4 E}
\r)\nonumber\\
&-&Cos^4 \theta_{13} Sin^2 2 \theta_{12} Sin^2 \l(\frac{\Delta_s
L}{4 E} \r)
\label{pee}
\end{eqnarray}
where  $\Delta_s=
(m_2^2-m_1^2) = 7.3 \times 10^{-5} eV^2$ is the solar mass scale, 
$\Delta_a = (m_3^2 -m_2^2) \simeq (m_3^2-m_1^2)= 2.5 \times 10^{-3} eV^2$
is the
atmsopheric mass scale, $Sin^2 2 \theta_{23}=1$ and $Sin^2
\theta_{12}=0.863$. We have taken the best fit values of these parameters
and ignored the errors as we want to illustrate the effect of $Sin^2
\theta_{13}$ on the survival probability and since the main source of
error is the statistical error in the experiment because of small number
of events. 
The spectrum of geo-neutrino events detetected can be expressed by 
  \begin{equation}
\frac{dN}{dE}=N_p~ t~ \sigma(E)\sum_X \l(\frac{dn_\nu }{dE}\r)_X
{\int_0}^{2R_\oplus} dL~ P_{ee}(E,L) \frac{n_X(L)}{\tau_x}
\end{equation}
where $N_p$ is the number of protons in detector fiducial volume,
$t$ the exposure time, $\sigma(E)$ the cross section of the
reaction $\bar \nu_e+p \rightarrow e^+ +n$ \cite{sigma}. Since the
threshold of the reaction is $E_0=m_n+m_e-m_p=1.804 MeV$, this
reaction detects neutrinos from the decay chain of $^{232}Th$
($E_{max}=2.25 MeV$) and $^{238}U$ ($E_{max}=3.26 MeV$), whereas
neutrinos from $^{40}K$ ($E_{max}=1.31 MeV$) are below threshold.
Kamiokande reports  events as a function of the visible energy
$E_{vis}=E-E_0 + 2 m_e \simeq E-0.8 MeV$ which is the energy
released on the annihilation of the positron by the ambient
electrons.$(dn_\nu /dE)_X$ represents the number of neutrinos per
energy interval produced by the decay of $X=Th,U$ \cite{spectrum}.
The number density $n_X(L)$ of $U,Th$ as a function of distance
$L$ from the detector has to be put in assuming some specific
earth model. We have taken the distribution given for
$n_X(r)/\tau_X$ from Tables VIII and X of reference
\cite{ref} for the Kamiokande detector. For the Himalayan detector we take
the crustal values of $n_U(r)/\tau_U$ from Figure 1 of reference
\cite{ref}. The $n/\tau$ values for Thorium is determined by taking the
same
ratio for $(n/\tau)_{Th}/(n/\tau)_U$ as the crustal distribution 
for Kamioka.  

The results for geo-neutrino events spectrum at Kamioka are shown in
Figure 1 for different values of $Sin^2 2 \theta_{13}$. The geo-neutrino
events spectrum for a possible detector \cite{ino} at the Himalayas is
shown in
Figure 2.
 We find that when we change $Sin^2 2 \theta_{13} $ from $0$ to $0.2$ the
oscillation probability averaged over the energy spectrum of
geo-neutrinos, $\langle P_{ee} \rangle$,  
decreases by $12 \%$. To achieve a $12 \%$ experimental accuracy in the
statistics
a total of about 100 events are needed. This can be achied in Kamiokande
with an exposure  of $ 3 \times 10^{32}$ proton-years. In the Himalayas
the geo-neutrino flux is higher since the Tibetan plateu is twice the 
thickness
of the average continental crust ($30 km$). As a result the geoneutrino
flux at the Himalayas is larger by by a factor of $1.8$ compared to the
flux at Kamioka. The same exposure at Himalayas will result is $187$
events and a statistical uncertainity of $7.3 \%$ which means that $Sin^2
2 \theta_{13} $ at the Himalayas can be probed down to $ 0.15$ for an
exposure
of $ 3 \times 10^{32}$ proton-years.             

If a geoneutrino detector can have an exposure of $16 \times 10^{32}$
proton 
years
( $5- kiloton$ detector run for $4$ years) then the statistical error is
down
to $5 \%$ and if the uncertainities in the BSE model \cite{bse} can also
be brought down to $5\%$ 
then there exists the possibility that one can probe $Sin^2 \theta_{13}$
down to $0.06$ which close to what can be achieved with proposed reactor
neutrino experiments \cite{white} (assuming that uncertainity in  other
neutrino
parameters -mainly $Sin^2 \theta_{12}$ can be brought to below $5\%$ level 
as well from other experiments).

\begin{figure}
 \label{Fig.1 }
\centering
\includegraphics[width=20cm]{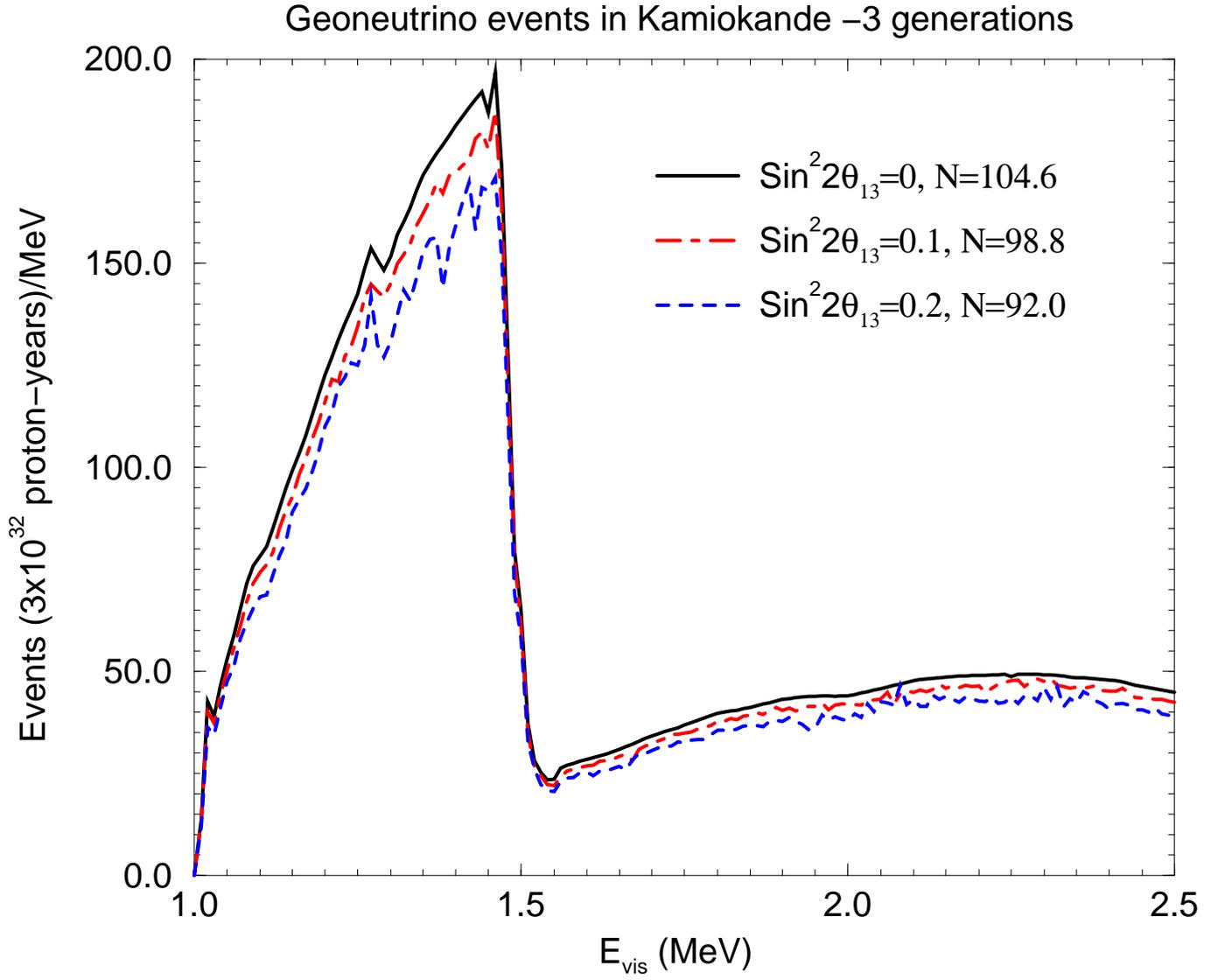}
\caption{Geoneutrino event spectrum and total number of events $N$
at Kamioka for different values of $Sin^2 2 \theta_{13}$. } \end{figure}

\begin{figure}
\label{Fig.2}
\centering
\includegraphics[width=20cm]{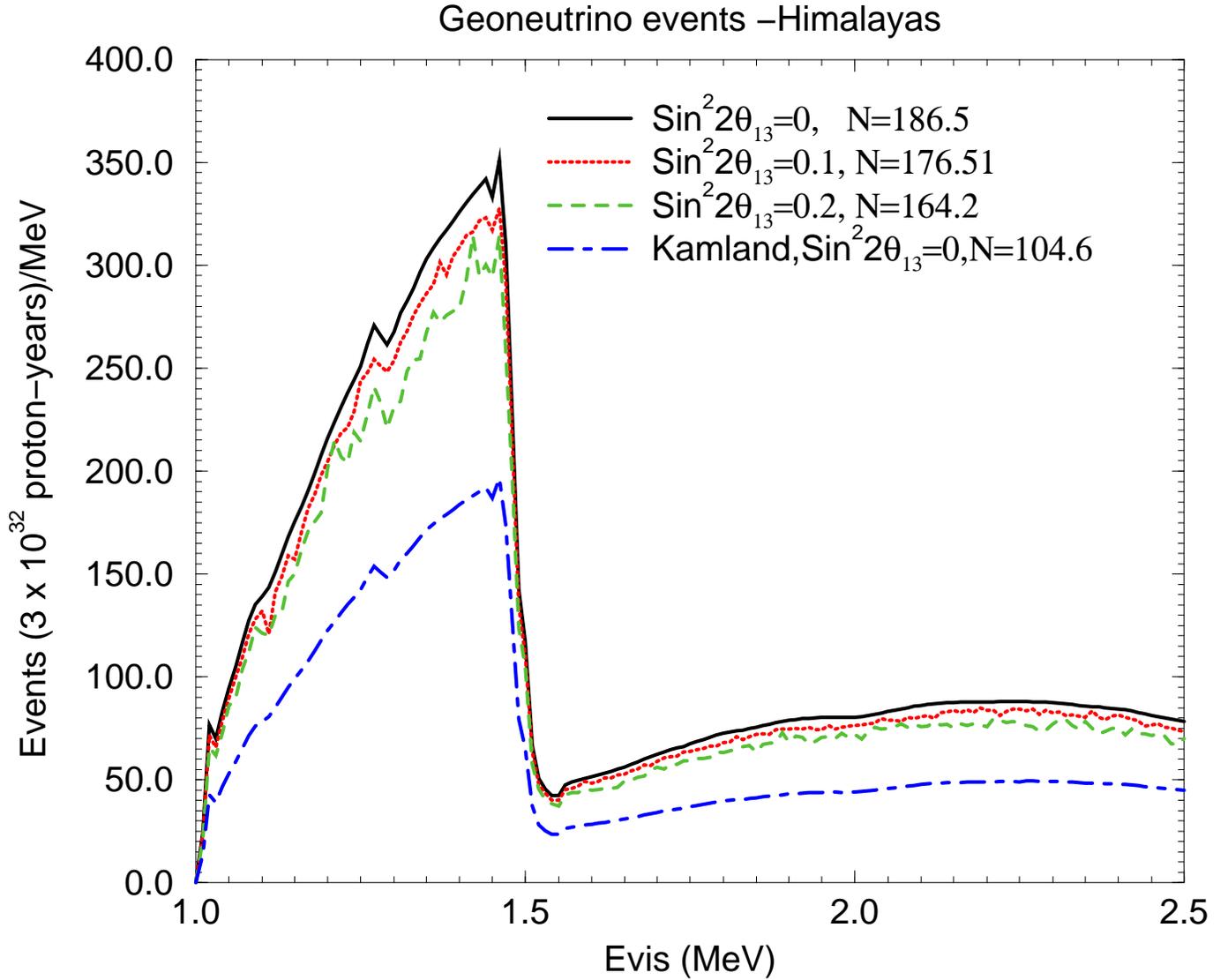}
\caption{Geoneutrino event spectrum and total number of events $N$
for a detector \cite{ino} close to Himalayas for different values of
$Sin^2 2
\theta_{13}$. The event spectrum at Kamioka is also shown for
comparison. } \end{figure}

\newpage

\end{document}